\documentclass[11pt]{article}
\usepackage[english]{babel}
\input{epsf}
\usepackage{mathtext}
\usepackage{graphics,dcolumn,bm,float}
\usepackage{amssymb,amsmath,rotate,color}
\usepackage{times}
\newcommand{\hhh}{{\cal H}}
\newcommand{\ii}{{\cal I}}

\newcommand{\CE}{{\cal E}}
\newcommand{\be}{\begin{equation}}
\newcommand{\ene}{\end{equation}}
\newcommand{\ba}{\begin{array}}
\newcommand{\ea}{\end{array}}

\newcommand{\bsigma}{\mbox{\boldmath$\sigma$}}

\begin{document}

\title{Dominant Majorana bound energy and critical current enhancement in ferromagnetic-superconducting topological insulator}
\author{M. Khezerlou\footnote{m.khezerlou@urmia.ac.ir}, H. Goudarzi\footnote{Corresponding author, e-mail address: h.goudarzi@urmia.ac.ir}, and S. Asgarifar\\
\footnotesize\textit{Department of Physics, Faculty of Science, Urmia University, P.O.Box: 165, Urmia, Iran}}
\date{}
\maketitle

\begin{abstract}
Among the potential applications of topological insulators, we theoretically study the coexistence of proximity-induced ferromagnetic and superconducting orders in the surface states of a 3-dimensional topological insulator. The superconducting electron-hole excitations can be significantly affected by the magnetic order induced by a ferromagnet. In one hand, the surface state of the topological insulator, protected by the time-reversal symmetry, creates a spin-triplet and, on the other hand, magnetic order causes to renormalize the effective superconducting gap. We find Majorana mode energy along the ferromagnet/superconductor interface to sensitively depend on the magnitude of magnetization $m_{zfs}$ from superconductor region, and its slope around perpendicular incidence is steep with very low dependency on $m_{zfs}$. The superconducting effective gap is renormalized by a factor $\eta(m_{zfs})$, and Andreev bound state in ferromagnet-superconductor/ferromagnet/ferromagnet-superconductor (FS/F/FS) Josephson junction is more sensitive to the magnitude of magnetizations of FS and F regions. In particular, we show that the presence of $m_{zfs}$ has a noticeable impact on the gap opening in Andreev bound state, which occurs in finite angle of incidence. This directly results in zero-energy Andreev state being dominant. By introducing the proper form of corresponding Dirac spinors for FS electron-hole states, we find that via the inclusion of $m_{zfs}$, the Josephson supercurrent is enhanced and exhibits almost abrupt crossover curve, featuring the dominant zero-energy Majorana bound states.
\end{abstract}

\section{INTRODUCTION}

Three-dimensional topological insulator (3DTI), which has been predicted theoretically \cite{KM} and discovered experimentally \cite{BZ,HK,QZ,HZ,KW} is characterized by gapless surface states and represents fully insulating gap in the bulk. Particularly, coincidence of the conduction and valence bands to each other in Dirac point, description of fermionic excitations as massless two-dimensional chiral Dirac fermions in the first Brillouin zone, depending chirality on the spin of electron, having the significant electron-phonon scattering on the surface and owning very low room-temperature electron mobility are the peculiar properties of electronic structure of the 3DTI. Due to spin-orbit interaction, the surface states are protected by the time-reversal symmetry, which are robust against perturbations. The spin of two-dimensional chiral Dirac-like charge carriers is tied to the momentum direction. It is highly desirable to investigate proximity-induced superconducting and ferromagnetic orders on the surface of a 3DTI \cite{FK,LTY,NA,ANB,FKT}. The broken spin rotation symmetry of the chiral surface states creates the spin-singlet component in topological insulator from conventional $s$-wave superconductor \cite{FK,SLT,JA,ORO}. On the other hand, the long-rang proximity effect in a conventional superconductor ferromagnet hybrid \cite{BVE} features the exotic odd-frequency spin-triplet component, which is odd under the exchange of time coordinates and even in momentum. This new superconducting condensation can be induced to the topologically conserved systems \cite{CBT}.
As a remarkable point, observing Majorana fermions, which have been detected in neutral systems, e.g. $Sr_2RuO_4$ and $3He$ $\cite{RG,MM,TKM}$, is of experimentally importance \cite{A,B}. In this regard, the interplay between a ferromagnet and superconductor on top of a 3DTI actually makes a sense to engineer the chiral Majorana mode  $\cite{LTY,TYN}$. Therefore, magnetic order can directly influence the transport manifestation in the superconducting 3DTI.

The search for transport properties of different hybrid structures including Majorana fermions has led to publish impressive number of guiding theoretical studies for experimental measurements \cite{FK,LTY,NA,TYN,F,SVG,LTYY}. In these systems, magnetization from a ferromagnet placed on top of the 3DTI can drastically affect the Andreev bound states. For instance, the magnetic order causes no $0-\pi$ oscillations in the critical current, leading to anomalous current-phase relation $\cite{LTY}$, in contrast to the metallic topologically trivial similar systems \cite{BBP}. Also, the direction of magnetization seriously influences the Josephson current, such that in-plane component of magnetization creates an intermediate phase shift, i.e. see Refs. \cite{LTY,TYN}. Recently, Burset et.al. \cite{BLT} have proposed normal/superconductor hybrid system deposited on top of the 3DTI, where the whole junction is exposed to a uniform magnetic order. However, it can be of particular interest that the magnetization induction to the superconducting and ferromagnetic regions in a N/S junction may be separately applied with a different magnitude for each region.

Regarding several works in the recent few years relating to the topological insulator-based junctions \cite{SVG,LTYY,BLT,LLN,NSB,VK,KG,FM,YZ,GK,ZY}, which are related to the Andreev process and resulting current-phase relation, we proceed, in this paper, to theoretically study the dynamical properties of Dirac-like charge carriers in the surface states of the 3DTI under influence of both superconducting and ferromagnetic orders. We introduce a proper form of corresponding Dirac spinors, which are principally distinct from those in Ref. \cite{BLT}.
The magnetization induction opens a gap at the Dirac point (no inducing any finite center of mass momentum to the Cooper pair), whereas the superconducting correlations creates an energy gap at the Fermi level, which is related to the chemical potential $\mu$. It is particularly interesting to investigate the topological insulator superconducting electron-hole excitation in the presence of a magnetic order. We assume that the Fermi level is close to the Dirac point, and the ferromagnet has a magnetization $|\textbf{M}|\leq\mu$. In the presence of magnetization, the chirality conservation of charge carriers in the surface states (due to opening the band gap in Dirac point) allows to use a finite magnitude of $|\textbf{M}|$. In similar systems without topological insulator, the spin-splitting arising from magnetization gives rise to limiting the magnitude of $|\textbf{M}|$ in a FS hybrid structure \cite{CC,CH}.
This excitation, therefore, is found to play a crucial role in Andreev process leading to the formation of Andreev bound state (ABS) between two superconducting segments separated by a weak-link ferromagnetic insulator. Particularly, we pay attention to the formation of Majorana bound mode as an interesting feature of the topological insulator F/S interface. We present, in section 2, the explicit signature of magnetization in low-energy effective Dirac-Bogogliubov-de Gennes (DBdG) Hamiltonian. The electron(hole) quasiparticle dispersion energy is analytically calculated, which seems to exhibit qualitatively distinct behavior in hole excitations ($|k_{fs}|<k_F$) by varying the magnitude of magnetization. By considering the magnetization being a finite value less than chemical potential in FS region, the corresponding proper eigenstates are analytically derived. Section 3 is devoted to unveil the above key point of FS correlations and respective discussion in a proposed FS/F/FS Josephson junction. In the last section, the main characteristics of proposed structure are summarized.

\section{THEORETICAL FORMALISM}
\subsection{Topological insulator FS effective Hamiltonian}

In order to investigate how both superconductivity and ferromagnetism induction to the surface state affects the electron-hole excitations in a 3DTI hybrid structure, we consider magnetization contribution to the DBdG equation.
Let us focus first on the Hubbard model Hamiltonian $\cite{HU}$ that is included the effective exchange field $\mathbf{M}$ follows from:
\begin{equation}
\hhh=-\sum_{\rho\rho^{'}s}t_{\rho\rho^{'}}\hat{c}^{\dagger}_{\rho s}\hat{c}_{\rho^{'}s}+\frac{1}{2}\sum_{\rho\rho^{'}ss^{'}}U_{\rho\rho^{'}ss^{'}}\hat{n}_{\rho s}\hat{n}_{\rho^{'}s^{'}}+\sum_{\rho ss^{'}}\hat{c}^{\dagger}_{\rho s}(\bsigma\cdot\mathbf{M})\hat{c}_{\rho s^{'}},
\end{equation}
where $U_{\rho\rho^{'}ss^{'}}$ denotes the effective attractive interaction between arbitrary electrons, labeled by the integer $\rho$ and $\rho^{'}$ with spins $s$ and $s^{'}$. The matrices $t_{\rho\rho^{'}}$ are responsible for the hopping between different neighboring sites, and $\hat{c}_{\rho s}$ and $\hat{n}_{\rho s}$ indicate the second quantized fermion and number operators, respectively. Here, $\bsigma (\sigma_x,\sigma_y,\sigma_z)$ is the vector of Pauli matrix. Using the Hartree-Fock-Gorkov approximation and Bogoliubov-Valatin transformation $\cite{HB}$, the Bogoliubov-de Gennes Hamiltonian describing dynamics of Bogoliubov quasiparticles is found. In Nambu basis, that electron(hole) state is given by $\Psi=\left(\psi_{\uparrow},\psi_{\downarrow},\psi^{\dagger}_{\uparrow},\psi^{\dagger}_{\downarrow}\right)$, the BdG Hamiltonian for a $s$-wave spin singlet superconducting gap in the presence of an exchange splitting can be written as:  
\begin{equation}
\hhh_{SF}=\left(\begin{array}{cc}
h(\mathbf{k})+M&\Delta(\mathbf{k})\\
-\Delta^{\ast}(-\mathbf{k})&-h^{\ast}(-\mathbf{k})-M
\end{array}\right),
\label{2}
\end{equation}
where $h(\mathbf{k})$ denotes the non-superconducting Schrodinger-type part, and $\Delta(\mathbf{k})$ is superconducting order parameter. In the simplest model, $\Delta(\mathbf{k})$ can be chosen to be real to describe time-reversed states. The effective exchange field by rotating our spin reference frame can be gain as $\left|\mathbf{M}\right|=\sqrt{m^{2}_{x}+m^{2}_{y}+m^{2}_{z}}$. The four corresponding levels of a singlet superconductor in a spin magnetic field is obtained $E_{s}(\mathbf{k})=\sqrt{\epsilon^{2}_{\mathbf{k}}+\left|\Delta(\mathbf{k})\right|^{2}}+s \left|\mathbf{M}\right|$ with $s=\pm 1$, where $\epsilon_{\mathbf{k}}$ is the normal state energy for $h(\mathbf{k})$. However, dependence of superconducting order parameter on the exchange energy can be exactly derived from self-consistency condition $\cite{PAG}$:
\begin{equation}
\Delta(\mathbf{k})=-\frac{1}{4}\sum_{\mathbf{k}s}U_{s-s}(\mathbf{k})\frac{\Delta_0(\mathbf{k})}{\sqrt{\epsilon^{2}_{\mathbf{k}}+\left|\Delta_0(\mathbf{k})\right|^{2}}}\tanh\left(\frac{\sqrt{\epsilon^{2}_{\mathbf{k}}+\left|\Delta_0(\mathbf{k})\right|^{2}}+s \left|\mathbf{M}\right|}{2k_BT}\right),
\label{a}
\end{equation}
where $\Delta_0(\mathbf{k})$ is the conventional order parameter in absence of ferromagnetic effect, $k_B$ and $T$ are the Boltzmann constant and temperature, respectively. The exchange splitting dependence of superconducting gap indicates that equation \eqref{a} has no functionality of $\mathbf{M}$ at zero temperature. This takes place under an important condition known as Clogston-Chandrasekhar limiting \cite{CC,CH}. According to this condition, if the exchange splitting becomes greater than a critical value $\left|\mathbf{M}_c\right|=\left|\Delta(T=0)\right|/\sqrt{2}$, then the normal state has a lower energy than the superconducting state. This means that a  phase transition from the superconducting to normal states is possible when the exchange splitting is increased at zero temperature.

We now proceed to treat such a ferromagnetic superconductivity coexistence at the Dirac point of a 3DTI. It should be stressed that the dressed Dirac fermions with an exchange field in topologically conserved surface state have to be in superconducting state. Here, the influence of exchange field interacts in a fundamentally different way comparing to the conventional topologically trivial system, where the exchange field splits the energy bands of the majority and minority spins. A strong TI is a material that the conducting surface states at an odd number of Dirac points in the Brillouin zone close the insulating bulk gap unless time-reversal symmetry is broken. Candidate Dirac-type materials include the semiconducting alloy $Bi_{1-x}Sb_x$, as well as HgTe and $\alpha-Sn$ under uniaxial strain \cite{STI}. In the simplest case, there is a single Dirac point in the surface Fermi circle and general effective Hamiltonian is modeled as $h^{TI}_{N}=\hbar v_{F}(\bsigma\cdot\mathbf{k})-\mu$,
where $v_{F}$ indicates the surface Fermi velocity, and $\mu$ is the chemical potential. Under the influence of a ferromagnetic proximity effect, the Hamiltonian for the two-dimensional surface states of a 3DTI reads as:
$$
h^{TI}_{F}=\hbar v_{F}(\bsigma\cdot\mathbf{k})-\mu+\mathbf{M}\cdot\bsigma,
$$
where the ferromagnetic contribution corresponds to an exchange field $\mathbf{M}=(m_{x},m_{y},m_{z})$. It has been shown $\cite{LTY}$ that transverse components of the magnetization on the surface $(m_{x},m_{y})$ are responsible to shift the position of the Fermi surface of band dispersion, while its perpendicular component to the surface induces an energy gap between conduction and valence bands.

In what follows, we will employ the relativistic generalization of BdG Hamiltonian, which is interacted by the effective exchange field to obtain the dispersion relation of FS dressed Dirac electrons in a topological insulator:
\begin{equation}
\hhh^{TI}_{FS}=\left(\begin{array}{cc}
h^{TI}_{F}(\mathbf{k})&\Delta(\mathbf{k})\\
-\Delta^{\ast}(-\mathbf{k})&-h^{TI\ast}_{F}(-\mathbf{k})
\end{array}\right).
\label{3}
\end{equation}
The superconducting order parameter now depends on both spin and momentum symmetry of the Cooper pair, that the gap matrix for spin-singlet can be given as $\Delta(\mathbf{k})=i\Delta_{0}\sigma_{y}e^{i\varphi}$,
where $\Delta_{0}$ is the uniform amplitude of the superconducting gap and phase $\varphi$ guarantees the globally broken $U(1)$ symmetry. By diagonalizing this Hamiltonian we arrive at an energy-momentum quartic equation. Without lose of essential physics, we suppose the component of magnetization vector along the transport direction to be zero $m_{x}=0$ for simplicity. Also, we set $m_{y}=0$, since the analytical calculations become unwieldy otherwise. The dispersion relation resulted from Eq. \eqref{3} for electron-hole excitations  is found to be of the form:
\begin{equation}
\CE_{FS}=\zeta\sqrt{\left(-\tau\mu_{fs}+\sqrt{m^{2}_{zfs}+\left|\mathbf{k}_{FS}\right|^{2}+\left|\Delta_0\right|^{2}(\frac{m_{zfs}}{\mu_{fs}})^{2}}\right)^{2}+\left|\Delta_0\right|^{2}\left(1-(\frac{m_{zfs}}{\mu_{fs}})^{2}\right)},
\label{5}
\end{equation}
where, the parameter $\zeta=\pm 1$ denotes the electron-like and hole-like excitations, while $\tau=\pm 1$ distinguishes the conduction and valence bands. We might expect several anomalous properties from the above superconducting excitations, which is investigated in detail in the next section. Equation \eqref{5} is clearly reduced to the standard eigenvalues for superconductor topological insulator in the absence of exchange field as $m_{z}=0$ (see Ref. \cite{LTY}), $\CE_{S}=\zeta\sqrt{\left(-\tau\mu_s+\left|\mathbf{k}_{S}\right|\right)^2+\left|\Delta_0\right|^{2}}$. The mean-field conditions are satisfied as long as $\Delta_{0}\ll\mu_{fs}$. In this condition, the exact form of superconducting wavevector of charge carriers can be acquired from the eigenstates $k_{fs}=\sqrt{\mu^2_{fs}-m^2_{zfs}}$.

The Hamiltonian Eq. \eqref{3} can be solved to obtain the electron (hole) eigenstates for FS topological insulator. The wavefunctions including a contribution of both electron-like and hole-like quasiparticles are analytically found as:
\begin{equation}
\psi^{e}_{FS}=\left(\begin{array}{cc}
e^{i\beta}\\
e^{i\beta}e^{i\theta_{fs}}\\
-e^{i\theta_{fs}}e^{-i\gamma_{e}}e^{-i\varphi}\\
e^{-i\gamma_{e}}e^{-i\varphi}
\end{array}\right)e^{i(k_{fs}^{x}x+k_{fs}^{y}y)}, \ \ \psi^{h}_{FS}=\left(\begin{array}{cc}
1\\
-e^{-i\theta_{fs}}\\
e^{i\beta}e^{-i\theta_{fs}}e^{-i\gamma_{h}}e^{-i\varphi}\\
e^{i\beta}e^{-i\gamma_{h}}e^{-i\varphi}
\end{array}\right)e^{i(-k_{fs}^{x}x+k_{fs}^{y}y)},
\end{equation}
where we define 
$$
\cos{\beta}=\frac{\CE_{FS}}{\eta\left|\Delta_0\right|} \ ; \ \ \eta=\sqrt{1-(\frac{m_{zfs}}{\mu_{fs}})^{2}} \ , \ \ e^{i\gamma_{e(h)}}=\frac{\Delta(\mathbf{k})}{\left|\Delta(\mathbf{k})\right|}.
$$
Note that, the solution is allowed as long as the Zeeman field being lower than chemical potential $m_{zfs}\leq \mu_{fs}$.

\subsection{FS/F/FS Josephson junction}

In what follows, we consider a line ferromagnetic junction of width $L$ between two FS sections in the coordinate ranges $x<0$ and $x>L$ in contact with TI surface states. The ferromagnetic region length is assumed to be smaller than the superconducting coherence length. The magnetizations of two superconducting semi-finite regions are taken to be equal and the same direction, see Fig. 1.
Coupling between electron and hole wave functions at the interface leads to scattering matrix, and it is necessary to solve the system of equations at each interface. We look for the energy spectrum for the ABSs, which can be obtain from a nontrivial solution for the boundary conditions $\Psi_{FS}^L=\Psi_F$ at $x=0$ and $\Psi_F=\Psi_{FS}^R$ at $x=L$, where the eigenvectors $\psi$ can be found in Appendix A. Solving the system of equations concerning to 8 electron-hole reflected and transmitted amplitudes coefficients results in a $8\times 8$-scattering matrix:
\begin{equation}
\mathcal{S}\Xi=0 \ ;\ \ \ \Xi=\left[t^e_L,t^h_L,a,b,c,d,t^e_R,t^h_R\right].
\end{equation}
The analytical expression of scattering matrix $\mathcal{S}$ is introduced in Appendix A.
The phase difference $\Delta\varphi=\varphi_R-\varphi_L$ is introduced by assuming that the phase of the left and right superconducting regions is $\varphi_L$ and $\varphi_R$, respectively. When determinant of $\mathcal{S}$ vanishes, then the nontrivial solution leads to an analytical expression for ABS in term of macroscopic phase difference $\Delta\varphi$
\begin{equation}
\epsilon(\Delta\varphi)=\eta\Delta_0\sqrt{\frac{1}{2}\left(1-\frac{\Gamma(\Delta\varphi)}{2\Omega}\right)} ,
\label{99}
\end{equation}
where we have defined
$$
\Gamma(\Delta\varphi)=\kappa_1\cos{\Delta\varphi}+\kappa_2\sin^{2}{(k^x_f L)} , \ \Omega=\kappa_3+\kappa_4\cos{(2k^x_f L)}.
$$
The explicit form of $\kappa_i (i=1,2,3,4)$ is given in Appendix A.
The analytical progress can be made in the limit of thin and strong barrier, which the barrier strength parameter is then defined as $Z=k_f^xL$. Otherwise, the results of analytical calculations can be presented as a function of length of junction. In short junction limit, the length of the junction is smaller than the superconducting coherence length $\xi=\hbar v_{F}/\Delta_0$.

We now analyze the supercurrent flowing through the Josephson junction. To calculate the normalized Josephson current in the short junction case which carried mostly by the ABS, the standard expression is introduced: 
\begin{equation}
\ii(\Delta\varphi)=\ii_{0}\int^{\pi/2}_{-\pi/2}\:d\theta_{fs}\cos{\theta_{fs}}\tanh{\left(\frac{\epsilon(\Delta\varphi)}{2K_{B}T}\right)}\frac{d\epsilon(\Delta\varphi)}{d\Delta\varphi},
\end{equation}
where $\ii_{0}=\left(e\left|\mathbf{k}_{FS}\right|W\Delta_0\right)/\left(\pi\hbar\right)$ is the normal current in a sheet of TI of width $W$. Note that the critical current can be measured as $\ii_c=\ii_0 max(\ii(\Delta\varphi))$. 
In the framework of such model, the ferromagnetic quasiparticles incidence angle may be real. The conservation of transverse component of the wave vectors allows us to find the propagation angle of electron or hole in the middle region, which is obtained to be of the following form:
\begin{equation}
\theta_f=\arcsin\left[\sqrt{\frac{\mu^2_{fs}-m^2_{zfs}}{\mu^2_f-m^2_{zf}}}\sin{\theta_{fs}}\right].
\label{t}
\end{equation}
It is worth noting the Eq. \eqref{t} indicates that the chemical potential in FS region may be lower than its magnitude in F region ($\mu_{f}>\mu_{fs}$). 

\section{RESULTS AND DISCUSSION}
\subsection{Energy excitation and Majorana mode}

In this section, we proceed to analyze in detail the dynamical features of Dirac-like charge carriers in 3DTI with ferromagnetic and superconducting orders deposited on top of it. We assume that the Fermi level controlled by the chemical potential $\mu$ is close to the Dirac point. In this case, it is expected the signature of $m_{zfs}<\mu_{fs}$ to be significant. In Fig. 2, we demonstrate the FS 3DTI electron-hole excitations. A net superconducting gap $\Delta_0$ is obtained in Dirac points (for $|k_{fs}|=k_F$, where $k_F$ is Fermi wavevector) when we set $m_{zfs}=0$. Increasing $m_{zfs}$ up to its possible maximum value results in three outcomes: i) the superconducting excitations, which is renormalized by a factor $|\Delta(\textbf{k})|\sqrt{1-(m_{zfs}/\mu_{fs})^2}$, disappear in hole branch ($|k_{fs}|<k_F$). It means that for the greater magnetizations, if we consent the superconductivity in FS 3DTI still exists, there is almost vanishing quantum state for reflected hole by Andreev process in the valence band, ii) Dirac point is shifted towards smaller FS quasiparticle electron-hole wavevectors, iii) the superconducting gap decreases slowly, where the variation of net gap is very low $\delta\Delta_0\ll|\Delta(\textbf{k})|$. The Andreev process, therefore, is believed to inconsiderably suppress. The signature of these valence band excitations can be clearly shown in AR, and as well in ABS, where the Majorana mode may also be formed at the 3DTI F/FS interface \cite{FK,TYN}.

As a verified result, considering the topological insulator interface between the ferromagnetic insulator and conventional superconductor leads to the appearance of the chiral Majorana mode as an Andreev bound state. In other words, the Majorana mode and Andreev reflection are strongly related to each other. The latter can be realized by the fact of looking for bound energies produced by the perfect AR, which yields the following solution:
\begin{equation}
\tilde{\epsilon}(\theta)=\eta\Delta_0 sgn\left(\Lambda\right)/\sqrt{1+\Lambda^2} \ ; \ \ \ \Lambda=\tan\left[\frac{1}{2i}\ln(\frac{\upsilon_1}{\upsilon_2})\right],
\end{equation}
where we define
$$
\upsilon_{1(2)}=4i\sin{k^{xe}_{f}L}\cos{\theta}\mathcal{M}_{2(1)}\mathcal{A}_{1(2)}+2e^{-ik^{xe}_{f}L}\cos{\theta}\mathcal{A}_{1(2)}-\mathcal{B}_{2(1)}\mathcal{A}_{1(2)}.
$$
The axiliarly parameters $\mathcal{M}_{2(1)}, \mathcal{A}_{1(2)}$ and $\mathcal{B}_{2(1)}$ are given in Apendix B.
We have checked numerically that sign of $\Lambda$ is changed by $sgn(m_{zf})$. Thus, the sign of Andreev resonance states may be changed by reversing the direction of $m_{zf}$, and it corresponds to the chirality of Majorana mode energies. As shown in Fig. 3, the slope of the curve of $\tilde{\epsilon}(\theta)$ around $\tilde{\epsilon}(\theta=0)=0$ shows no change with the increase of $m_{zfs}/\mu_n$ for fixed $m_{zf}$, while it exhibits significantly decreasing behavior with the increase of $m_{zf}/\mu_n$ for fixed $m_{zfs}$. The dispersion of Majorana modes along the interface ($\theta=\pi/2$) decreases with the increase of both magnetizations of FS and F regions. Note that, due to the presence of $m_{zfs}$ it needs to consider the Fermi level mismatch between normal and FS sections, i.e. $\mu_n\neq\mu_{fs}$. Then, the above contributions can be considerable in Andreev process and resulting Josephson supercurrent.

\subsection{Current-phase relation}

In order to study the influence of ferromagnetic order in FS region on the supercurrent passing through the FS 3DTI Josephson junction, we proceed to focus on the ABS. As a usual result in similar systems, the $4\pi$ periodic gapless bound energy corresponding to chiral Majorana bound is obtained in normal incidence $\theta_{fs}=0$ for finite magnitude of $m_{zfs}<\mu_s$. In sharp contrast to the previous results \cite{LTY}, which increase of the magnetization magnitude of F region $m_{zf}$ causes to flattening the ABS, the opening of the gap in finite angles of incidence presented in $\Delta\varphi=\pm\pi$ is suppressed with the increase of the $m_{zfs}$. Also, the minimum of gaped ABSs comes through the higher angle of incidence. These are demonstrated in Figs. 4(a) and (b). According to Majorana mode of Sec. 3.1, the gapless Andreev Majorana bound along the interface decreases with the increase of $m_{zfs}$, e.g. it reaches a value $0.6\tilde{\epsilon}_{max}(\Delta\varphi)$ for $m_{zfs}=0.8\mu_s$.

However, the chirality of Majorana ABS, which is observed in Figs. 3 and 4, can be provided by the fact that the FS 3DTI Hamiltonian solutions may be protected by time-reversal symmetry. It is now easy to see that the wavefunctions Eq. (6) $\psi_{FS}^e$ and $\psi_{FS}^h$ are connected by the time-reversal operation as
$$
\psi_{FS}^h(\theta_{fs}=0)=\tau_0i\sigma_y\psi_{FS}^{e*}(\theta_{fs}=0),
$$
where $\tau_0$ denotes a unit matrix in Nambu space. Moreover, $\psi_{FS}^e(\theta_{fs}=0)$ and $\psi_{FS}^h(\theta_{fs}=0)$ are the eigenstates of the helicity matrix $\tau_z\sigma_x$ satisfying the eigenvalue equation
$$
\tau_z\sigma_x\psi_{FS}^{e(h)}(\theta_{fs}=0)=\pm\psi_{FS}^{e(h)}(\theta_{fs}=0),
$$
and orthogonal to each other. They, therefore, immune to spin-independent potential scattering and, hence, the ferromagnetic exchange field $m_{zfs}$ can play an important role in transport properties.

In what follows, we consider the angle-averaged supercurrent originated from Andreev bound states in which the zero-temperature limit is assumed in the following plots. The $2\pi$ periodic current-phase curve is found for different values of $m_{zfs}$, as shown in Fig. 4(c), with a shape far from sinusoidal. One can say that almost abrupt crossover curve is seen from Josephson current originating from the zero energy states (ZES), whereas the supercurrent peak does not appear in maximum phase difference $\Delta\varphi=\pi$. The latter originates from the presence of ferromagnetic-proximity in topological insulator. Consequently, in the presence of high magnetization of FS region (e.g. $m_{zfs}=0.8\mu_s$) and low magnetization of F region ($m_{zf}<0.2\mu_n$), the Majorana bound energy may dominate the formation of angle-averaged Josephson current. This is in agreement with the Majorana bound energy, shown in Fig. 4(b).
Finally, to see the effect of $m_{zfs}$ on the critical current, we plot the width $L/\xi$ of F region dependence of critical current, where $\xi$ is the superconducting coherence length. Increasing the $m_{zfs}$ causes to increase the maximum of supercurrent. The critical current exhibits oscillatory function in terms of length of junction, where the oscillation-amplitude is considerable and decreases with the decrease of $m_{zfs}$.

\section{CONCLUSION}

In summary, we have investigated the influence of ferromagnetic and superconducting orders proximity at the same time on the surface state of a topological insulator. The superconducting topological insulator electron-hole excitations in the presence of magnetization led to achieve qualitatively distinct transport properties in the FS/F/FS Josephson junction. It has been shown that for magnetic order $m_{zfs}\approx\mu_s$, the spin-triplet component becomes dominant. One of key findings of the present work is the appearance of novel Majorana bound mode at the F/FS interface, which can be controlled by the tuning of magnetization magnitude of FS region. The $4\pi$ periodic gapless Andreev bound states, corresponding to the Majorana bound energy, were created for finite magnitude of $m_{zfs}\leq\mu_s$. The current-phase relation curve has been found to be far from sinusoidal, and its critical value shows increasing with the increase of the magnetization of FS region. Note that, these results have been obtained in the case, when $m_{zfs}, m_{zf}\leq\mu$ and $\mu\gg\Delta_0$, which is relevant to the experimental regime.

\renewcommand{\theequation}{A-\arabic{equation}}
  \setcounter{equation}{0}  
  \section*{APPENDIX A: Josephson scattering matrix}  
By introducing the scattering coefficients in F region we write down the total wave function inside the F region:  
$$
\Psi_{F}=e^{ik^{y}_{f}y}\left(a\psi^{e+}_Fe^{ik^{xe}_{f}x}+b\psi^{e-}_Fe^{-ik^{xe}_{f}x}+c\psi^{h+}_Fe^{-i k^{xh}_{f}x}+d\psi^{h-}_Fe^{i k^{xh}_{f}x}\right),
$$
where right and left moving electron and hole spinors can be written as:
$$
\psi^{e+}_{F}=\left[1,\alpha e^{i\theta_f},0,0\right]^{T},\ \psi^{e-}_{F}=\left[1,-\alpha e^{-i\theta_f},0,0\right]^{T},
$$
$$
\psi^{h+}_{F}=\left[0,0,1,\alpha e^{i\theta_f}\right]^{T}, \ \psi^{h-}_{F}=\left[0,0,1,-\alpha e^{-i\theta_f}\right]^{T}.
$$

Also, in this appendix, we characterize the 8$\times$8 matrix $\mathcal{S}$ in the form of four 4$\times$4 matrices which is used to calculate the Andreev energy bound states and corresponding Josephson supercurrent in the FS/F/FS junction:
$$
\mathcal{S}=\left(\begin{array}{cc}
\mathcal{S}_1&\mathcal{S}_{2}\\
\mathcal{S}_{3}&\mathcal{S}_{4}
\end{array}\right),
$$
where
$$
\mathcal{S}_1=\left(\begin{array}{cccc}
e^{i\beta}&e^{-i\beta}&-1&-1\\
-e^{i\beta}e^{-i\theta_{fs}}&e^{-i\beta}e^{i\theta_{fs}}&-\alpha e^{i\theta_f}&\alpha e^{-i\theta_f}\\
e^{-i\theta_{fs}}e^{-i\varphi_L}&-e^{i\theta_{fs}}e^{-i\varphi_L}&0&0\\
e^{-i\varphi_L}&e^{-i\varphi_L}&0&0
\end{array}\right);\ \ \ \mathcal{S}_2=\left(\begin{array}{cccc}
0&0&0&0\\
0&0&0&0\\
-1&-1&0&0\\
-\alpha e^{i\theta_f}&\alpha e^{-i\theta_f}&0&0
\end{array}\right);
$$
$$
\mathcal{S}_3=\left(\begin{array}{cccc}
0&0&-e^{ik^{xe}_{f}L}&-e^{-ik^{xe}_{f}L}\\
0&0&-\alpha e^{i\theta_f}e^{ik^{xe}_{f}L}&\alpha e^{-i\theta}e^{-ik^{xe}_{f}L}\\
0&0&0&0\\
0&0&0&0
\end{array}\right);
$$
$$
\mathcal{S}_4=\left(\begin{array}{cccc}
0&0&e^{i\beta}e^{ik^{x}_{fs}L}&-e^{-i\beta}e^{-ik^{x}_{fs}L}\\
0&0&e^{i\beta}e^{i\theta_{fs}}e^{ik^{x}_{fs}L}&-e^{-i\beta}e^{-i\theta_{fs}}e^{-ik^{x}_{fs}L}\\
-e^{-ik^{xh}_{f}L}&-e^{ik^{xh}_{f}L}&-e^{i\theta_{fs}}e^{-i\varphi_R}e^{ik^{x}_{fs}L}&e^{-i\theta_{fs}}e^{-i\varphi_R}e^{-ik^{x}_{fs}L}\\
-\alpha e^{i\theta_f}e^{-ik^{xh}_{f}L}&\alpha e^{-i\theta_f}e^{ik^{xh}_{f}L}&e^{-i\varphi_2}e^{ik^{x}_{fs}L}&e^{-i\varphi_2}e^{-ik^{x}_{fs}L}
\end{array}\right).
$$

The definition of $\kappa_i$ quantities in Andreev bound state energy are given by:
$$
\kappa_1=-16\alpha^2 \cos^2{\theta_{fs}}\cos^2{\theta_f},
$$
$$
\kappa_2=16\alpha\sin{\theta_{fs}}\sin{\theta_f}(1+\alpha^2)-16\alpha^2+4\cos{2\theta_{fs}}(1+\alpha^4)+8\alpha^2\cos{2\theta_f},
$$
$$
\kappa_3=-4\alpha\sin{\theta_f}\sin{\theta_{fs}}(1+\alpha^2)+4\alpha^2+(1+\alpha^4)+2\alpha^2\cos{2\theta_f}\cos{2\theta_{fs}},
$$
$$
\kappa_4=4\alpha\sin{\theta_f}\sin{\theta_{fs}}(1+\alpha^2)-4\alpha^2\sin^2{\theta_{fs}}+2\alpha^2\cos{2\theta_f}-(1+\alpha^4).
$$

\renewcommand{\theequation}{B-\arabic{equation}}
  \setcounter{equation}{0}  
  \section*{APPENDIX B: Majorana mode energy}
The parameters related to the Majorana mode energy of Eq. (11) read as following:
$$
\mathcal{A}_1(2)=e^{(-)i\theta_{fs}}\left[(-)\mathcal{N}_1(\mathcal{M}_{2(1)}-1)e^{ik^{xe}_{f}L}-(+)\mathcal{N}_2\mathcal{M}_{2(1)}e^{-ik^{xe}_{f}L}\right],
$$
$$
\mathcal{B}_1(2)=-\mathcal{N}_1(\mathcal{M}_{1(2)}-1)e^{-ik^{xe}_{f}L}+\mathcal{N}_2\mathcal{M}_{1(2)}e^{ik^{xe}_{f}L},
$$
$$
\mathcal{N}_1(2)=(-)\alpha e^{(-)i\theta_f}+e^{-i\theta},\ \ \ \mathcal{M}_{1(2)}=\frac{\alpha e^{i\theta_f}-(+)e^{(-)i\theta_{fs}}}{2\alpha\cos{\theta_f}}.
$$

\newpage

\textbf{Figure captions}\\
\textbf{Figure 1} (color online) Sketch of the topological insulator-based FS/F/FS Josephson junction. The magnetization vectors in F and FS regions are prependicular to the surface of topological insulator.\\
\textbf{Figure 2} (color online) The ferromagnetic superconducting excitation spectra on the surface state of 3DTI for several values of $m_{zs}$, calculated from Eq. (5). We set the net value of superconducting gap $|\Delta_S|=0.5\;eV$ (this value of pair potential is taken only to more clarify the behavior of spectra in Dirac point, although it does not further need to use it in our calculations, since $\mu_{fs}/|\Delta_S|\gg 1$ is supposed.\\
\textbf{Figure 3} (color online) The dispersion of Majorana modes as a function of the electron incident angle for several values of magnetizations in FS and F regions. The solid lines correspond to $m_{zf}=0.2\mu_n$ and the dashed lines to $m_{zfs}=0.2\mu_n$.\\
\textbf{Figure 4(a), (b), (c)} (color online) Plot of the Andreev bound state energy versus phase difference and superconductor quasiparticle angle of incidence in Josephson FS/F/FS junction. The role of the magnetization of FS region $m_{zfs}$ is demonstrated. Curve (a) represents $m_{zfs}=0.1\mu_{fs}$ and (b) $m_{zfs}=0.8\mu_{fs}$. We set $\mu_{fs}=100\Delta_0$, $\mu_f=2\mu_{fs}$ and $m_{zf}=0.2\mu_{fs}$ and $L/\xi=0.05$. Plot (c) represents the normalized Josephson supercurrent as a function of the phase difference with respect to varying $m_{zfs}=0.8, 0.5, 0.1\mu_{fs}$ when $m_{zf}=0.4\mu_{fs}$. The critical current $J_c/J_0$ is plotted versus lenght of junction for two magnitudes of $m_{zfs}=0.8\mu_{fs}$ (solid line) and $0.2\mu_{fs}$ (dashed line).\\

\newpage

\begin{figure}[p]
\epsfxsize=0.3 \textwidth
\begin{center}
\epsfbox{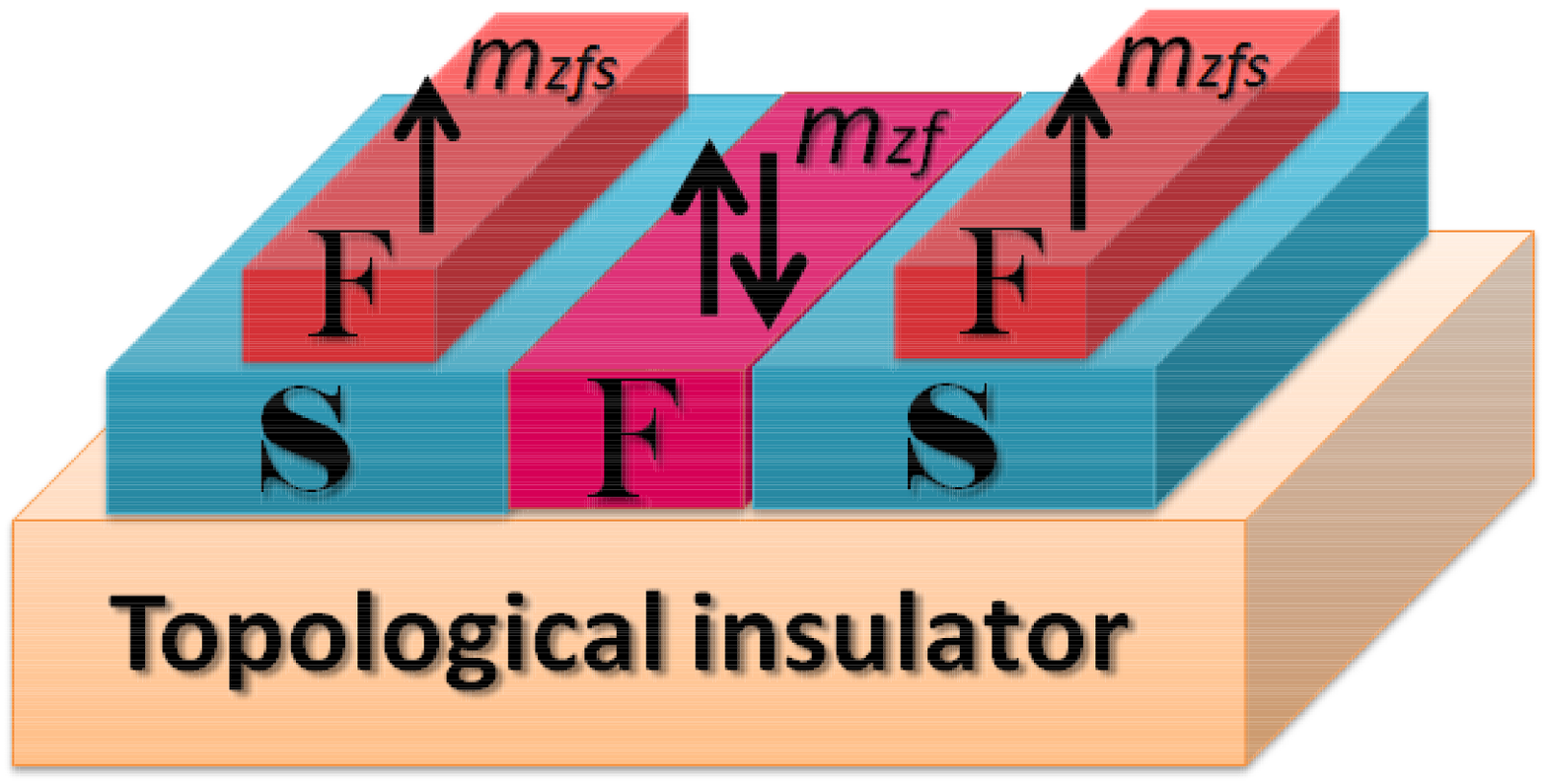}
\setcounter{figure}{0}
\caption{\footnotesize }
\end{center}
\end{figure}

\begin{figure}[p]
\epsfxsize=0.5 \textwidth
\begin{center}
\epsfbox{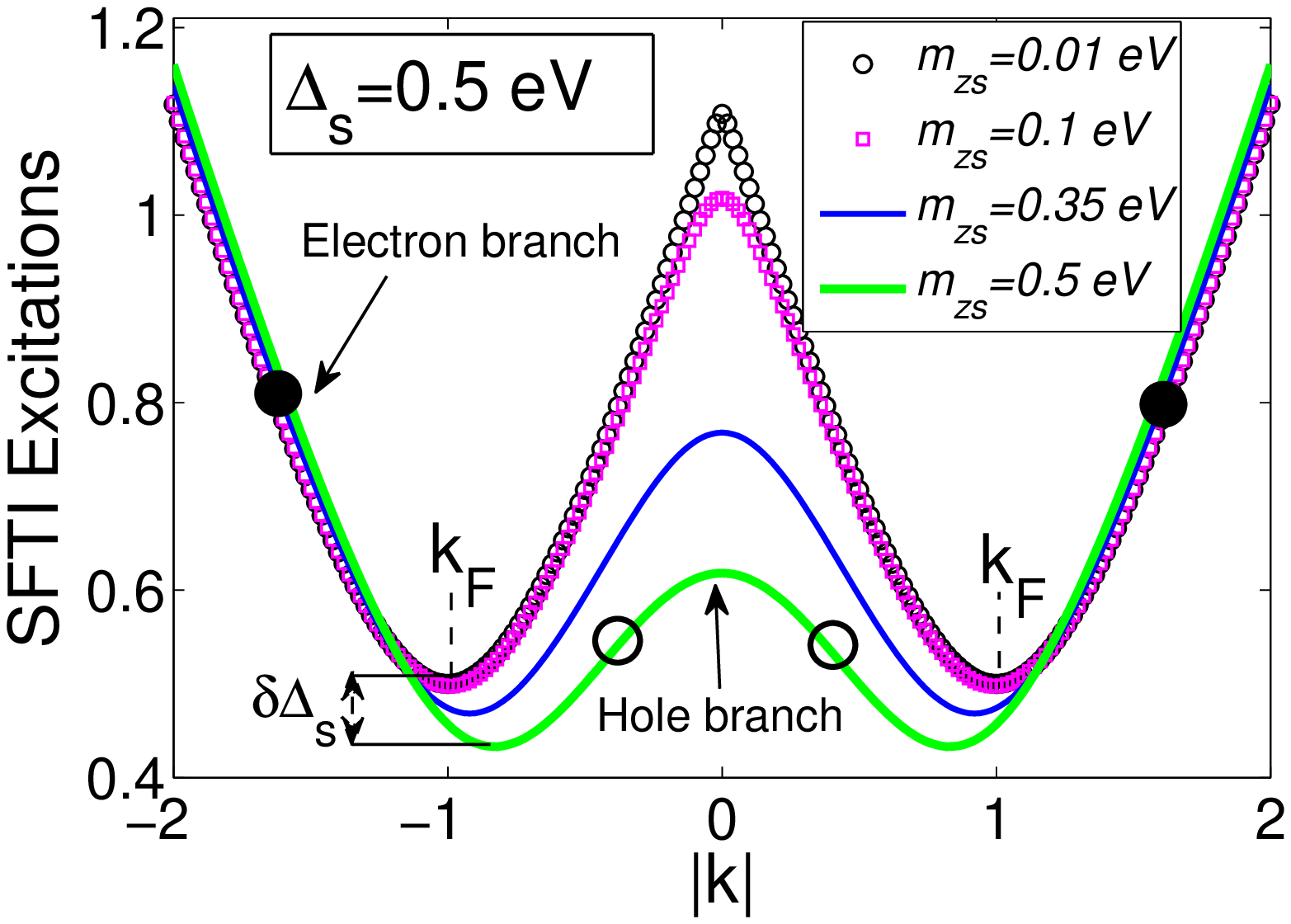}
\setcounter{figure}{1}
\caption{\footnotesize }
\end{center}
\end{figure}

\begin{figure}[p]
\epsfxsize=0.5 \textwidth
\begin{center}
\epsfbox{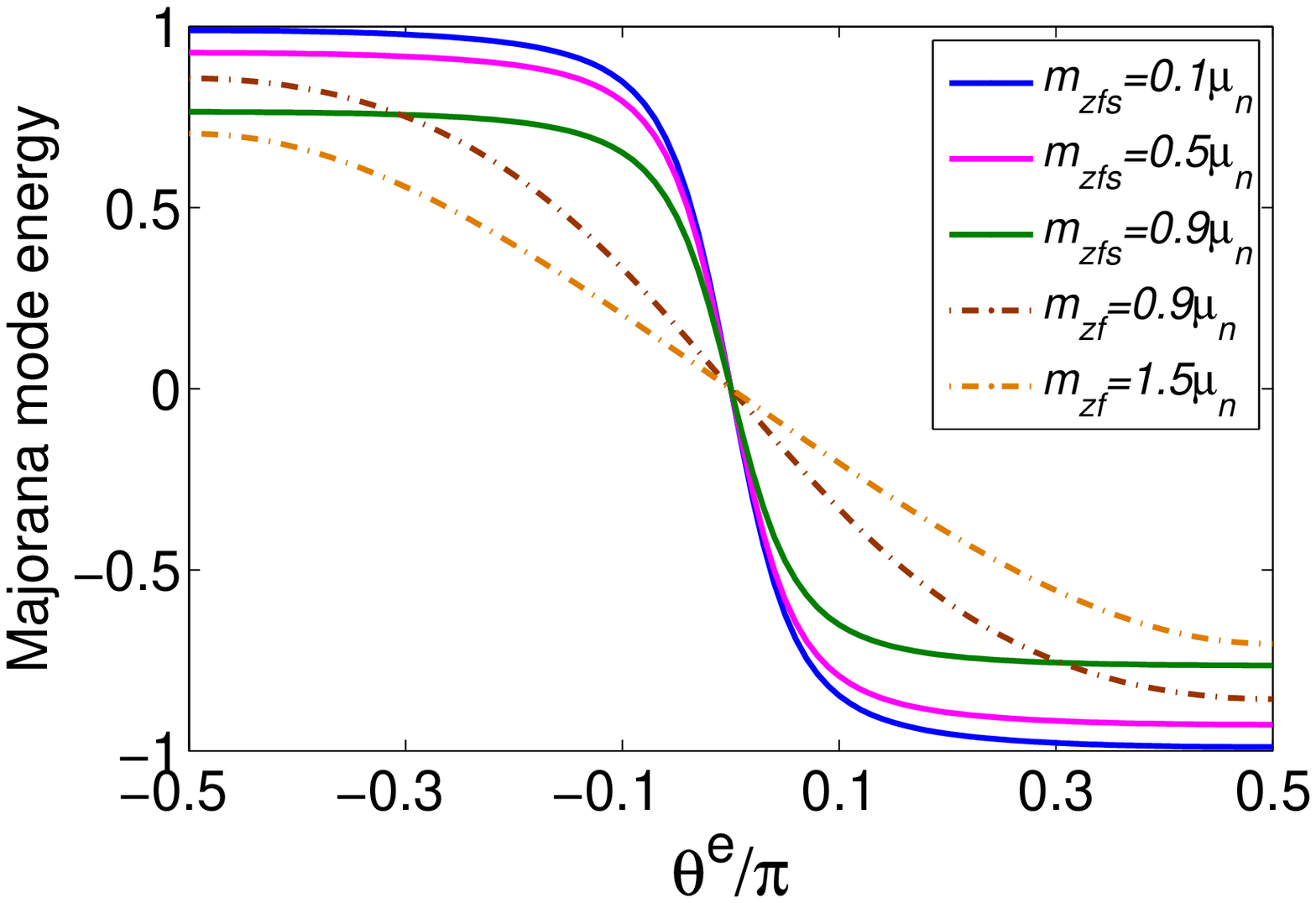}
\setcounter{figure}{2}
\caption{\footnotesize }
\end{center}
\end{figure}

\begin{figure}[p]
\epsfxsize=0.5 \textwidth
\begin{center}
\epsfbox{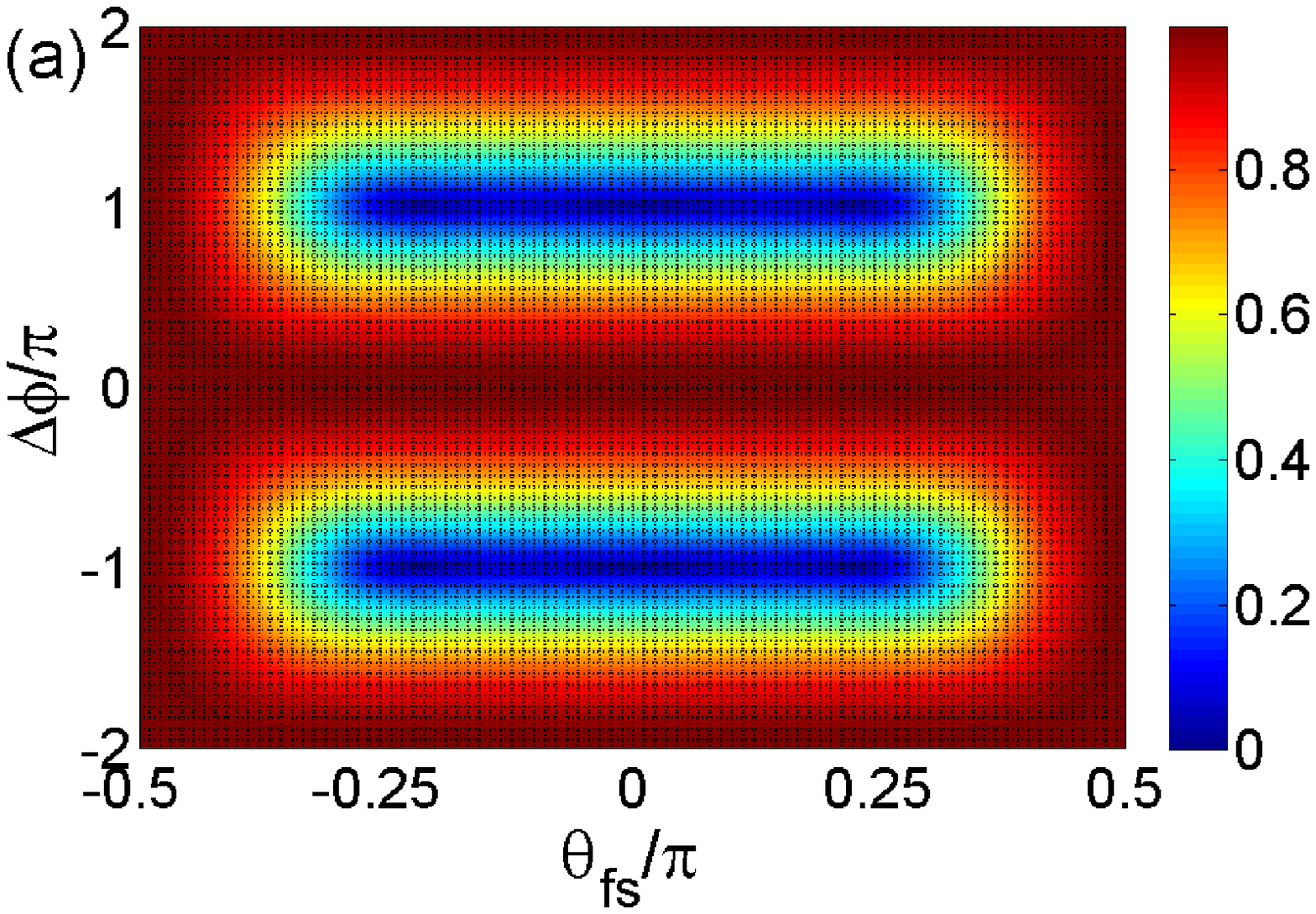}
\end{center}
\end{figure}

\begin{figure}[p]
\epsfxsize=0.5 \textwidth
\begin{center}
\epsfbox{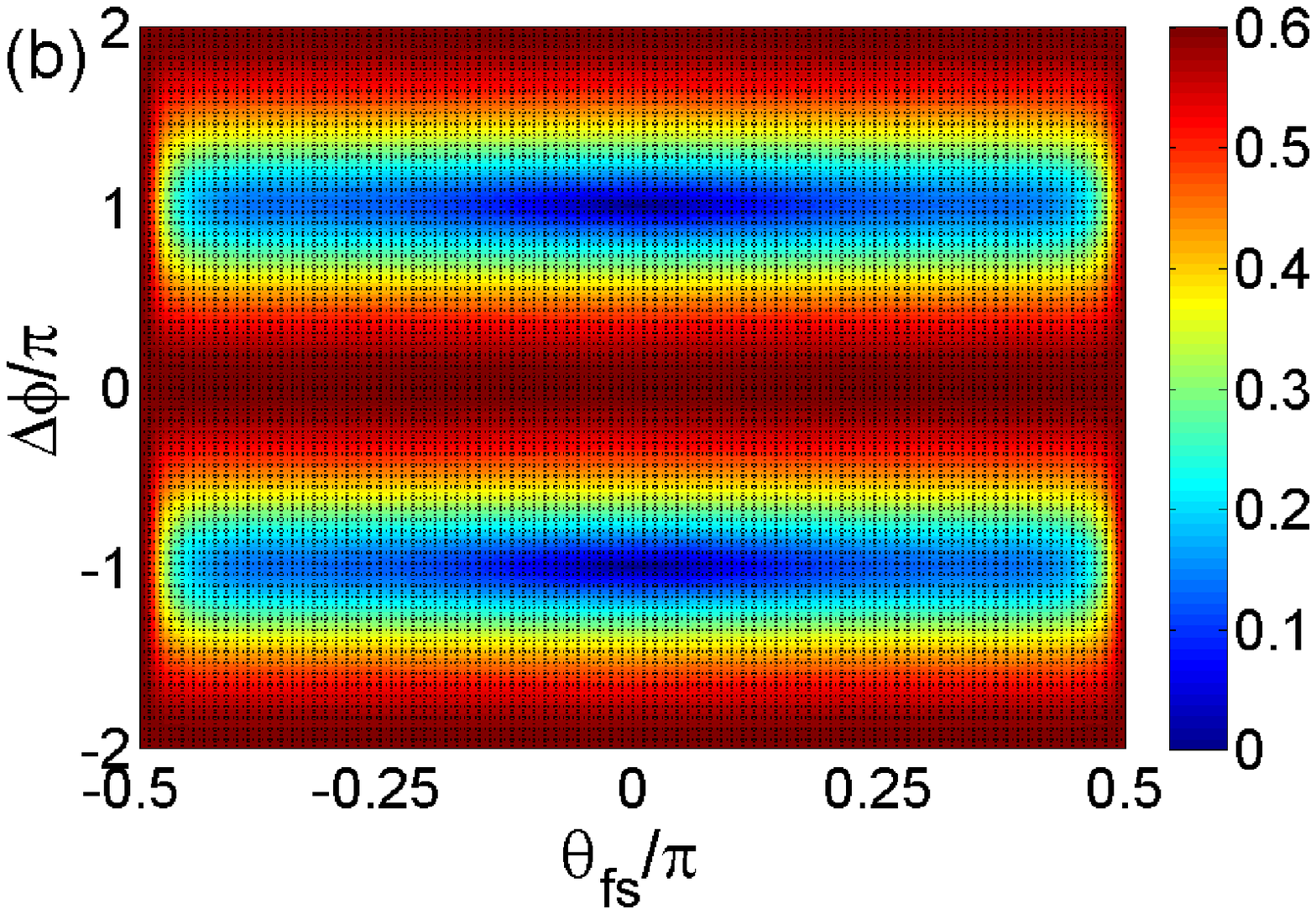}
\end{center}
\end{figure}

\begin{figure}[p]
\epsfxsize=0.5 \textwidth
\begin{center}
\epsfbox{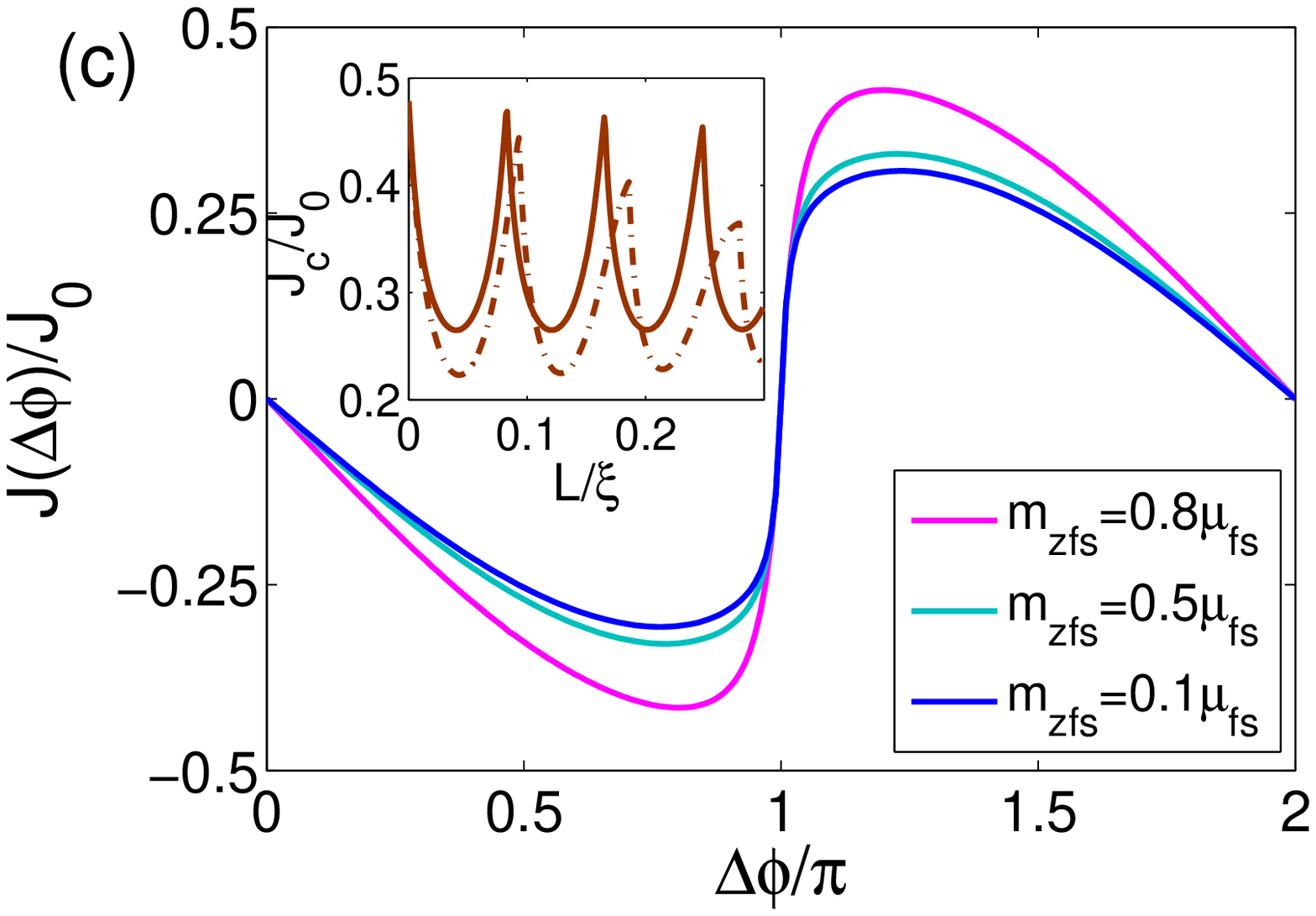}
\setcounter{figure}{3}
\caption{\footnotesize (a),(b),(c)}
\end{center}
\end{figure}

\end{document}